\documentclass[sigconf,authorversion]{acmart}
\AtBeginDocument{%
  \providecommand\BibTeX{{%
    \normalfont B\kern-0.5em{\scshape i\kern-0.25em b}\kern-0.8em\TeX}}}

\copyrightyear{2024}
\acmYear{2024}
\setcopyright{acmlicensed}\acmConference[MUM '24]{International Conference on Mobile and Ubiquitous Multimedia}{December 1--4, 2024}{Stockholm, Sweden}
\acmBooktitle{International Conference on Mobile and Ubiquitous Multimedia (MUM '24), December 1--4, 2024, Stockholm, Sweden}
\acmDOI{10.1145/3701571.3701604}
\acmISBN{979-8-4007-1283-8/24/12}

\usepackage{color, colortbl}
\usepackage{subfig}
\usepackage{multirow}
\usepackage{longtable}
\usepackage{hyperref}
\usepackage[font=small,labelfont=bf]{caption}
\usepackage{wasysym}
\usepackage[nolist,nohyperlinks, printonlyused, withpage]{acronym}
\usepackage{enumitem}
\usepackage{caption}
\usepackage{subcaption}

\begin{document}

% "Level Up or Game Over: The Impact of Dark Patterns on Mobile Game Ethics"
% "The Final Boss: Confronting Dark Patterns in Mobile Gaming"
% "Loot Boxes and Level Caps: What Drives Mobile Games to the Dark Side?"
% "Respawn in the Shadows: How Dark Patterns Reshape Mobile Gaming"
% "Quest for Ethics: Uncovering Dark Patterns in Mobile Games"
% "Grinding for Darkness: The Role of Unethical Design in Mobile Games"
% "Player vs. Dark Patterns: An Analysis of Unethical Game Mechanics"
% "Game Mechanics or Manipulation? The Rise of Dark Patterns in Mobile Gaming"
% "Dark Souls of Mobile Games: Navigating the Ethics of Dark Patterns"
% "From Bonus Rounds to Dark Routines: Exploring Unethical Design in Mobile Games"

\title[Level Up or Game Over]{Level Up or Game Over: Exploring How Dark Patterns Shape Mobile Games}

%The Real Boss Fight: Investigating Dark Patterns in Mobile Gaming

\author{Sam Niknejad}
\email{sam.niknejad@hotmail.de}
\orcid{0009-0004-9195-5917}
\authornote{Both authors contributed equally to the paper}
\affiliation{%
  \institution{University of Bremen}
  %\city{Bremen}
  \country{Germany}
}

\author{Thomas Mildner}
\email{mildner@uni-bremen.de}
\authornotemark[1]
\orcid{0000-0002-1712-0741}
\affiliation{
    \institution{Digital Media Lab\\University of Bremen}
    %\city{Bremen}
    \country{Germany}
}

\author{Nima Zargham}
\email{zargham@uni-bremen.de}
\orcid{0000-0003-4116-0601}
\affiliation{
  \institution{Digital Media Lab\\University of Bremen}
  %\city{Bremen}
  \country{Germany}
}

\author{Susanne Putze}
\email{sputze@uni-bremen.de}
\orcid{0000-0002-3072-235X}
\affiliation{%
  \institution{Digital Media Lab\\University of Bremen}
  %\city{Bremen}
  \country{Germany}
}

\author{Rainer Malaka}
\email{malaka@uni-bremen.de}
\orcid{0000-0001-6463-4828}
\affiliation{
    \institution{Digital Media Lab\\University of Bremen}
    %\city{Bremen}
    \country{Germany}
}

%%%%%%%%%%%%%%%%%%%%%
\renewcommand{\shortauthors}{Niknejad and Mildner et al.}

\begin{acronym}
\acro{HCI}{Human-Computer Interaction}
\end{acronym}

%%%%%%%%%%%%%%%%%%%%%%%%%%%%
%%%%%    "Abstract"    %%%%%
%%%%%%%%%%%%%%%%%%%%%%%%%%%%
\begin{abstract}
% The present study focuses on dark patterns in mobile games, which are deceptive design elements used to manipulate users. Relevant research in the field of dark patterns is considered, and established methods are employed to investigate dark patterns in mobile games. Using a custom-designed crawler, information about reported dark patterns in mobile games was collected and visualized in graphs. The study reveals the subtle nature and widespread usage of these manipulation techniques, even in games thought to be healthy. The findings suggest that future research should consider incorporating user-generated content and community engagement to enhance awareness of dark patterns in game development, promoting ethical design practices.
This study explores the prevalence of dark patterns in mobile games that exploit players through temporal, monetary, social, and psychological means. Recognizing the ethical concerns and potential harm surrounding these manipulative strategies, we analyze user-generated data of 1496 games to identify relationships between the deployment of dark patterns within ``dark'' and ``healthy'' games. Our findings reveal that dark patterns are not only widespread in games typically seen as problematic but are also present in games that may be perceived as benign. This research contributes needed quantitative support to the broader understanding of dark patterns in games. With an emphasis on ethical design, our study highlights current problems of revenue models that can be particularly harmful to vulnerable populations. To this end, we discuss the relevance of community-based approaches to surface harmful design and the necessity for collaboration among players/users and practitioners to promote healthier gaming experiences.
\end{abstract}

%%%%%%%%%%%%%%%%%%%%%%%
%%%%%    "CCS"    %%%%%
%%%%%%%%%%%%%%%%%%%%%%%


%%%%%%%%%%%%%%%%%%%%%%%%%%%%
%%%%%    "Keywords"    %%%%%
%%%%%%%%%%%%%%%%%%%%%%%%%%%%

\begin{teaserfigure}
\centering
\includegraphics[width=1\columnwidth]{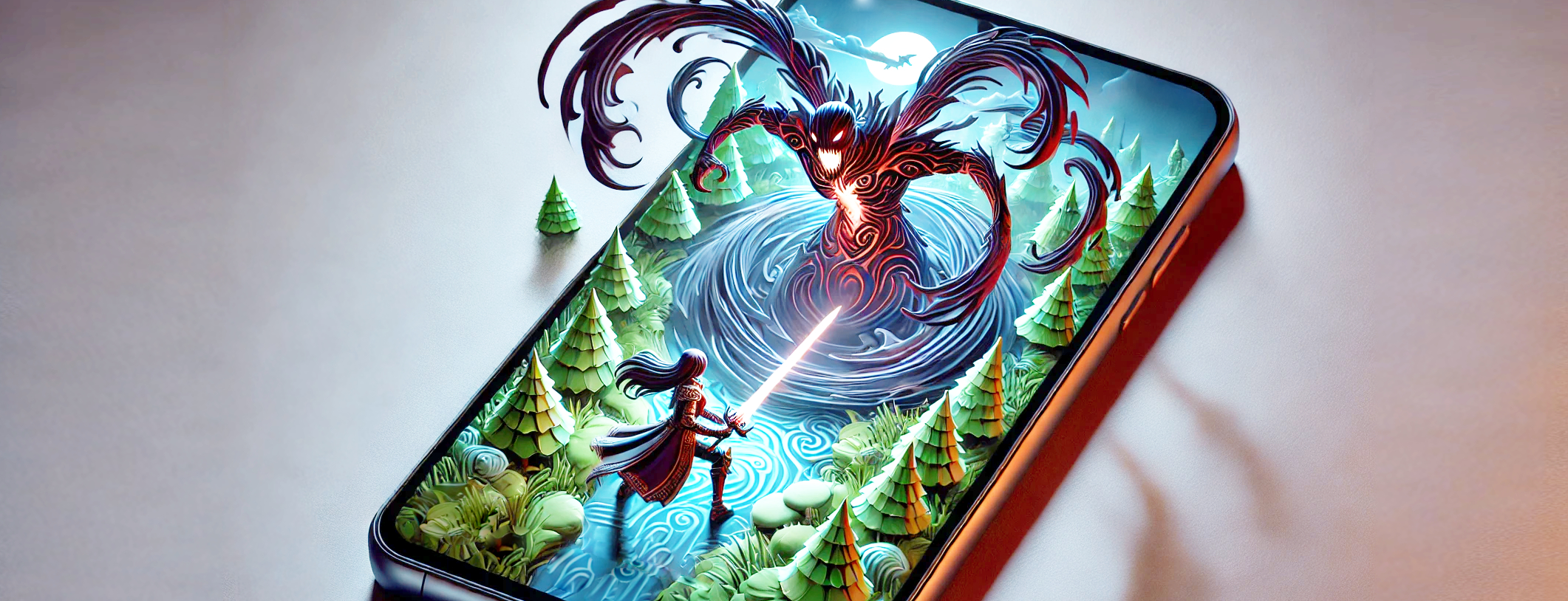}
\caption{AI generated image showing a mobile device on which a fighting scene between a heroine and a monster takes place to represent the a fight against dark patterns. The image was generated using \href{https://openai.com/index/dall-e-3/}{DALL-E 3}.}
% \description{AI generated image showing a mobile device on which a fighting scene between a heroine and a monster takes place to represent the a fight against dark patterns. The image was generated using DALL-E 3.}
\label{fig:teaser}
\end{teaserfigure}
%\textcolor{red}{\textbf{Draft: September 27, 2023}}

\maketitle
%%%%%%%%%%%%%%%%%%%%%%%%%%%%%
%        MOTIVATION/Forschungsfragen         %
%%%%%%%%%%%%%%%%%%%%%%%%%%%%%
\section{Introduction} 
%In 2022, the total revenue in the Games market is projected to reach $155.10 billion, with an anticipated compound annual growth rate (CAGR) of 8.42% from 2022 to 2027, resulting in a projected market volume of $249.90 billion by 2027. The revenue from in-app purchases (IAP) is expected to reach $75.06 billion, revenue from paid apps is projected at $1.16 billion, and advertising revenue is forecasted to be $78.84 billion in 2022. The market is anticipated to see 118.70 billion downloads, with an average revenue per download of $1.31\citep{statista_Games_Worldwide}.
The gaming industry is a steadily growing market with number of players steadily increasing~\cite{statista_number_2020} and more purchases recorded over the past years~\cite{statista_inapp_2020}. Today, the industry flourishes as e-sports and streaming services offer new opportunities for additional revenue, e.g., through advertisement~\cite{gawrysiak_using_2020}, further amplifying its growth. This is reflected by moves in the industry from selling full game titles to alternative payment models. As one example, free-to-play models have become more widespread, especially in mobile apps, providing players with essential gaming experiences that can be enhanced through in-app purchases~\cite{davidovici-nora_innovation_2013, cai_who_2019}.

Aside from certain benefits for the industry and its consumers, an ongoing discourse covers the potential negative implications of excessive gaming on players' well-being and health~\cite{kiraly_gaming_2023}.
In 2015, a cross-national study in European countries with a particular focus on internet gaming disorder revealed that $3\%$ of adolescents who regularly engage in gaming are affected, with a further $8.4\%$ being at risk~\cite{muller_regular_2015}. Generally, problems in self-control can be linked to unhealthy behavior~\cite{cudo_relationship_2020}.
%the German Center for Addiction Issues in Childhood and Adolescence noted a surge in pathological gaming behavior and social media addiction among children and adolescents during the pandemic, as reported in the German Medical Journal.
From an HCI and design perspective, recent efforts to capture and describe unethical designs -- specifically ``dark patterns''\footnote{We are aware of a recent choice by the ACM Diversity, Equity, and Inclusion Council to categorise the term ``dark pattern'' as problematic~\cite{acm_words_2023}. However, we opted to use the term to stay in line with related work describing unethical and harmful UI practices. We acknowledge that the term could be misinterpreted with a connotation that ``dark'' could refer to ``bad'' or ``evil' intentions. However, we argue that the term makes a reference to \textit{hidden} or \textit{obfuscated} consequences for users~\cite{brignull_deceptive_design}, where intentions play a secondary role.} -- provide some pointers directed at practices that negatively affect user experience by tricking users into unwanted behaviour and restricting available choices~\cite{brignull_deceptive_design, Gray2018}. Dark patterns have been shown to influence users into making undesirable choices~\cite{gray_ontology_2024} with instances described in e-commerce~\cite{mathur_dark_2019}, social media~\cite{mildner_about_2023, schaffner_understanding_2022}, and mobile applications~\cite{di_geronimo_ui_2020, Gray2018}. A critical issue of these unethical practices lies in the inability of users to effectively recognise and avoid their effects, as indicated in several studies (e.g., \cite{luguri_shining_2021, mildner_defending_2023, bongard-blanchy_i_2021, di_geronimo_ui_2020}).

%This trend is concerning, particularly due to the large number of young and adolescent players involved in gaming communities. The Office for Technology Assessment at the German Bundestag criticized dark pattern as "unethical, sometimes unfair, and potentially fraudulent" in November 2019. Given their extensive impact on privacy, consumer protection, and other areas, it is essential to research dark patterns to fully understand their effects and ethical implications.

\begin{figure*}[t!]
    \centering
    \includegraphics[width=.55\textwidth]{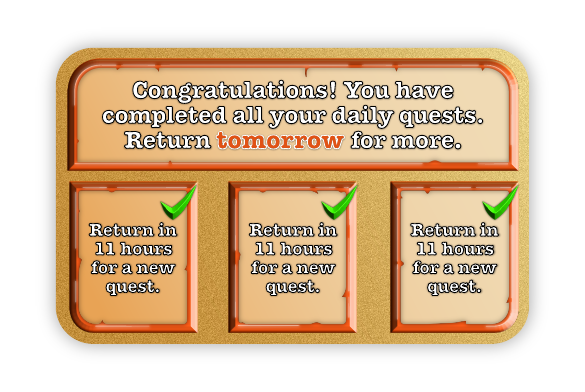}
    \caption{This figure presents a mock-up example based on Zagal et al.'s~\cite{zagal_dark_2013} ``playing by appointment'' dark pattern in the form of quests that require users to play the game on a daily basis. The displayed graphic is the original work of one of this paper's authors and does not violate any licensing.}
    \label{fig:playing-by-appointment}
    \Description{
    This figure demonstrates a mock example of a playing-by-appointment dark pattern by Zagal et al. that offers players daily achievements but requires regular logins and activities. The latter forces players to play the game daily in order to get the most out of it and not get a disadvantage.
    }
\end{figure*}

While dark patterns have been explored and studied in various domains~\cite{mathur_what_2021, gray_mapping_2023, chang_theorizing_2024}, there remains a research gap, particularly in gaming contexts. In this scope, Zagal et al.~\cite{zagal_dark_2013} conceptualised domain-specific dark patterns typical to game-related interactions. In their work, the authors provide definitions for seven types within three categories: temporal, monetary, and social capital-based. Figure~\ref{fig:playing-by-appointment} includes an example of their temporal ``playing by appointment'' dark pattern, which forces players to return to a game daily in order to complete quests. Recently, Hadan et al.~\cite{hadan_motivating_2024} investigated dark patterns in games by building on Zagal et al.'s definitions following a qualitative approach, similar to other work on the topic (e.g. \cite{Gray2018},\cite{mildner_about_2023},\cite{gunawan_comparative_2021}). Although these works are needed to understand underlying mechanisms and users' perceptions of dark patterns, quantitative approaches help understand the range in which certain dark patterns occur. Mathur et al.'s~\cite{mathur_dark_2019} large-scale analysis of online shopping sites spearheaded these kinds of efforts. However, similar studies are relatively scarce. Following up on Zagal et al.'s discussion, which emphasises a misalignment between creators' interests and players' welfare, we aim to extend the exploration of dark patterns in games by presenting the results of a quantitative study utilising user-generated data.% from the online platform \url{darkpatterns.games}.

%We seek to augment the contemporary dark patterns discourse by leveraging user-generated data to provide empirical insights into the prevalence and characteristics of dark patterns in mobile games.
To this end, we explore a large set of user reviews based on the website \href{http://www.darkpatterns.games/}{\textit{www.darkpatterns.games}}~\cite{noauthor_darkpatterngames_nodate}, considering $1.496$ mobile iOS and Android games. The website adapts Zagal et al.'s~\cite{zagal_dark_2013} dark pattern categories and includes a fourth, thus differentiating between temporal, monetary, social, and psychological types of dark patterns. In line with Zagal et al's~\cite{zagal_dark_2013} work, temporal dark patterns refer to strategies prolonging or shortening certain interaction sequences against players' expectations. They manifest, for instance, as ``grinding'' or ``playing by appointment''. Monetary dark patterns concern instances which trick players into spending more money than they originally anticipated. ``Pay to skip'' describes mechanics which offer shortcuts to overcome otherwise cumbersome gameplay interactions (e.g. grinding). Social (capital-based) dark patterns refer to designs that exploit social aspects and incorporate strategies such as ``social pyramid schemes.'' Added by the authors of the \href{http://www.darkpatterns.games/}{\textit{www.darkpatterns.games}} website, psychological tricks are referenced as strategies deployed to misguide players into making bad choices~\cite{noauthor_darkpatterngames_nodate}, which are part of Zagal et al.'s discussion~\cite{zagal_dark_2013}. Building on the framework of dark pattern categories, our exploratory analysis addresses the following research question:

\begin{itemize}
        \item[\textbf{RQ:}] How prevalent are types of dark patterns in mobile games? %To what degree do categories of dark patterns manifest in mobile games?
\end{itemize}

Through our analysis, we offer quantitative support for existing dark pattern frameworks. Drawing on Zagal et al.'s work~\cite{zagal_dark_2013}, we demonstrate the effectiveness of their categories. By utilising large-scale user-generated data, we also provide novel insights into users' general ability to identify dark patterns and their important role in combating unethical design through community-based efforts. 

%By employing a statistical framework, we strive to elucidate patterns and trends within the data, thereby contributing to a more comprehensive understanding of the ethical challenges inherent in contemporary game design practices. Our objective is to provide actionable insights that can inform the development of ethical guidelines and best practices for game designers, ultimately fostering a gaming industry that prioritizes player well-being and operates with transparency and integrity.

%In light of these concerns, this paper addresses several guiding questions:
% In light of these concerns, this paper addresses several key guiding questions:
% \begin{enumerate}
% \item What differentiations exist within various dark pattern?
% \item When does a game qualify as Healthy, and when does it transition into Dark territory, and how do they relate to each other?
% \item Are Healthy Games completely devoid of dark pattern?
% \item Which category is the most popular and boasts a larger user base?
% \item Is payment always necessary to access dark pattern-free games?
% \item  How do the frequency and nature of dark pattern differ between the two datasets?
% \item  Are there specific dark pattern that are more prevalent in Dark Games compared to Healthy Games?
% \end{enumerate}

%%%%%%%%%%%%%%%%%%%%%%%%%%%%%
%%%%%   RELATED WORK    %%%%%
%%%%%%%%%%%%%%%%%%%%%%%%%%%%%
\section{Background \& Related Work}
This section outlines the related work this study builds on. First, we begin with a background on dark patterns research. Second, we discuss mobile games and player's experience with such games. Lastly, we follow research constituting the current dark pattern landscape in games.

%%%%%%%%%%%%%%%%%%%%%%
%%%%%%%%%%%%%%%%%%%%%%
\subsection{Research on Dark Patterns}
%%% Dark Pattern as a Term
Over a decade ago, Harry Brignull~\cite{brignull_deceptive_design} first coined the term ``dark pattern'' to describe interface tricks that coerce users' choices to their disadvantage. 
%%% What Are Dark Patterns
Following this effort, work addressing the same concept through slightly differing terminology -- including deceptive design patterns and damaging patters -- focus on misleading or coercive strategies that seek to divert the users' attention towards an end goal that is not in the user's best interest~\cite{gray_ontology_2024, mildner_about_2023, mathur_what_2021, monge_roffarello_defining_2023}.
%%%%%%%%%% Research on Dark Pattern is Expanding
Today, this body of research spans e-commerce and online shopping sites~\cite{mathur_dark_2019, Gray2018}, social media~\cite{mildner_about_2023, schaffner_understanding_2022}, and games~\cite{zagal_dark_2013, hadan_motivating_2024}. A recent publication describes a comprehensive ontology for dark patterns by Gray et al.~\cite{gray_ontology_2024}, synthesising these and further work on the topic in a granulated structure. In doing so, the ontology differentiates between high-, meso-, and low-level types of dark patterns, offering a shared vocabulary that affords easier communication for a field that gains complexity with more domains being investigated. While high- and meso-level types usually describe technology and domain-agnostic strategies, low-level types contain domain-specific dark patterns. Although Zagal et al.'s~\cite{zagal_dark_2013} dark patterns were not included in the initial ontology, they fit within the low-level category. As our study focuses on a specific domain in games, we will remain focused on the low-level types described in Zagal. et al.'s work.

%%%%%%%%%% Dark Patterns is Found in Different Domains

%%%%%%%%%% Dark Patterns in XR
%\citet{Hadan2024} conducted a systematic analysis of deceptive design research for extended reality (XR) applications. Authors found that XR's immersive capabilities and extensive data collection enable subtle and powerful manipulation strategies.

%%%%% Dark Design Mobile Apps
With regard to peoples' ability to recognise dark patterns, work has identified challenges among users to avoid their harmful effects~\cite{bongard-blanchy_i_2021,mildner_defending_2023}. Investigating the presence of dark patterns in mobile applications, \citet{di_geronimo_ui_2020} studied how users engage with them based on a cognitive walkthrough study. The authors considered $240$ apps across various categories, which they analyzed for the presence of any dark patterns, reporting that $95\%$ of these apps contained at least one instance. Their user study demonstrated that most participants were unable to detect dark patterns. Taking inspiration from this work, \citet{mildner_defending_2023} followed a similar approach to investigate social media users' ability to recognise dark patterns. Their work echoes prior findings that (1) users are capable of differentiating between interfaces with and without dark patterns on a statistical level but further implies (2) that this capability is insufficient to effectively avoid dark patterns' harms. By considering a larger-scale data collection from a community-based platform, our work adds further quantitative insights to these existing qualitative studies with respect to mobile games.

%%%%%%%%%%%%%%%%%%%%%%
%%%%%%%%%%%%%%%%%%%%%%
\subsection{Player Experience in Mobile Games}
%%%%%% Rise of games
While video games have been around for quite some time, they have matured into becoming one of the most popular forms of entertainment worldwide, engaging millions of players across various platforms~\cite{Tang2020}. The gaming landscape continuously evolves, offering players a wide range of experiences that cater to diverse interests and preferences~\cite{Zargham2024LetsTalk}. In recent years, the proliferation of smartphones and tablets has led to mainstream adoption of mobile games~\cite{Baabdullah2020Factors}. This specific form of gaming offers portability with a focus on shorter engagement periods~\cite{Bowman2015Observed} but also motivates continuous engagement and spending~\cite{Syvertsen2022Problem} with ethical implications.
%%% Mobile Game Characteristics
% Unlike computer or console games, mobile games offer portability, simple mechanics, greater emphasis on short events, and easy-to-learn repeatable gameplay experience, and in-app purchasing options \cite{Bowman2015Observed}.
%%% History
% One of the earliest mobile games designed is Snake, a simple puzzle game in which players steer a digital ``snake'' around walls and boxes while consuming an Item \cite{Bowman2015Observed}.
%%% More recent games
%Since their existence before smartphones, mobile games have since developed into a substantial and dynamic gaming market %world-renowned games such as Angry Birds \cite{angrybirds}, Among Us \cite{amongus}, and Candy Crush \cite{candycrush}, played by millions of players globally.

%%% Business Model Shift
%Overall, video game business models have undergone significant changes over the past few decades. 
%Supported by growing player-bases, business models behind the gaming industry have changed in recent years. 
Through the support of growing player bases, these changes have further affected business models used to monetize content.
Initially, game publishers generated income primarily through up-front purchases made via retail sales~\cite{King2023, van2021behavioral}. However, the rise of digital distribution, mobile gaming, and online platforms has introduced new revenue streams, particularly in the form of microtransactions. 
%%% Freemium Games
To this end, the ``freemium'' business model \cite{kerr2006business} stands behind various successful (mobile) games. The term translates to services that are free to download and use but offer in-app purchases that allow players to buy virtual goods, currency, or additional content. This also includes encouraging players to pay for advantages through better equipment or performance enhancers (as Zagal et al.'s ``monetized rivalries'' dark pattern~\cite{zagal_dark_2013} [p. 5]) or purchase game versions that add further functions to the games, such as getting rid of in-game advertising (like ``pre-delivered content''~\cite{zagal_dark_2013} [p. 5]).
%%% Artificial Obstacles
In a similar vein, work has shown~\cite{Syvertsen2022Problem} how often mobile games include structural characteristics that encourage players to make in-app purchases. These characteristics, referred to as ``artificial obstacles,'' are intentionally designed to slow game progression unless players invest additional time or money.
%%% Bridge and Connection to this work
In this work, we investigate which deceptive tactics are implemented in mobile games and to what extent players can identify different dark patterns in this context.

%%%%%%%%%%%%%%%%%%%%%%
%%%%%%%%%%%%%%%%%%%%%%
\subsection{Dark Patterns in Games}
To our knowledge, \citet{zagal_dark_2013} were among the first to explore dark patterns in gaming contexts. Their work categorizes dark patterns into three groups: temporal, monetary, and social capital-based dark patterns. Temporal dark patterns manipulate the time players must invest to progress, often leading to frustration and incentivizing spending to speed up gameplay. Monetary dark patterns focus on financial manipulation, such as enticing players to make frequent microtransactions or disguising the true cost of in-game purchases. Social capital-based dark patterns exploit players' social networks and relationships, pressuring them to recruit friends or compete with peers to advance in the game.
Since this work, additional research has built on this taxonomy. 
To raise awareness about the various types of monetization in freemium apps, \citet{Fitton2019}, for instance, developed the App Dark Design (ADD) framework. In this work, the authors highlight a particular concern regarding younger app users and the potential harms associated with different ``dark design'' aspects in mobile apps. Fitton et al. utilize Zagal et al.'s taxonomy to demonstrate the particular harm children, who describe a vulnerable population, face when playing free-to-play and thus easily accessible games.
Also building on Zagal et al.'s work, \citet{sousa_dark_2023} follow a game's transition from a buy-to-play to a free-to-play business model between versions. Using a qualitative approach, the work maps adopted deceptive strategies to Zagal et al.'s taxonomy, showing an increase since the business model changed. 

%%% Bridge and Connection to this work
%This work expands the research of dark patterns by examining deceptive design approaches and their various forms in mobile games. 
%%%% Zagal
% As noted earlier, Zagal et al. classified deceptive patterns in games into three categories: temporal, monetary, and social capital-based dark patterns~\cite{zagal_dark_2013}. Temporal dark patterns manipulate the time players must invest to progress, often leading to frustration and incentivizing spending to speed up gameplay. Monetary dark patterns focus on financial manipulation, such as enticing players to make frequent microtransactions or disguising the true cost of in-game purchases. Social capital-based dark patterns exploit players' social networks and relationships, pressuring them to recruit friends or compete with peers to advance in the game. Their taxonomy has been applied to a website~\footnote{\url{https://www.darkpattern.games}}, which hosts a crowd-sourced repository of deceptive pattern examples found in approximately 20,000 mobile games. 

%%% Predatory Monetization
Giving this kind of strategy a name, researchers use ``predatory monetization'' to refer to purchasing systems that hide the long-term costs from players until they are already financially and psychologically invested \cite{king2018predatory, petrovskaya2021predatory}. A prominent example is loot boxes, which are reward systems where players buy packages containing a random series of virtual items. By manipulating the probability of obtaining desirable items, designers encourage players to make repeated purchases. This approach has been associated with problem gambling \cite{Zendle2020}.
%%%% Dark Patterns in Games Part 1
\citet{Karlsen2019Exploited} investigated design patterns that encourage players to spend more time on games than they initially intended and how these patterns might contribute to compulsive and problematic gaming habits. The work suggests that while some design elements may make games tedious, others can significantly contribute to gaming addiction. 
%%%% Deceptive Design in Games
With the aim to explore the use of deceptive design for monetization within popular ``freemium'' 3D games, \citet{King2023} conducted an online survey with $259$ people. They identified six categories of deceptive design patterns within the chosen context: Predatory Monetization, Default to Purchase, UI Misdirection, Emotional Interpersonal Persuasion, Physical Placement, and Narrative Obligation.

%%%% Gap
Despite the widespread popularity of mobile games, research in this area remains limited compared to studies on non-mobile forms of gaming \cite{Syvertsen2022Problem}. This gap in research extends to the exploration of potential adverse consequences associated with mobile gaming, such as addiction, financial impacts due to in-app purchases, and the social implications of prolonged game use.
%%%% Our Contribution
This work aims to address this gap by exploring the prevalence and impact of dark patterns in mobile games through a quantitative approach. 
%shedding light on potential ethical considerations and implications for consumer welfare and protection.

%%%%%%%%%%%%%%%%%%%%%%%%%%%%%
%%%%     METHODOLOGY     %%%%
%%%%%%%%%%%%%%%%%%%%%%%%%%%%%
\section{Methodology}
To answer our research question and expand our current understanding of dark patterns in mobile games, we crawled the website \href{www.darkpattern.games}{\textit{www.darkpattern.games}} after informing the host and gaining their permission. The data only includes publicly available ratings for individual games, excluding users' personal information. At the time of conducting this experiment, in June 2024, the website listed a total of $52,111$ mobile games. Users of the website can rate each game based on three categories adopted from \citet{zagal_dark_2013}  -- temporal, monetary, social -- as well as a fourth to capture psychological dark patterns. Moreover, each category contains between seven and twelve variants (see Appendix \ref{app:dp-games-categories}). %Accumulating user ratings, 
When rating a game, users are asked in a binary way whether a dark pattern is present or not. Depending on overall results, the website sorts each game either into ``dark'' or ``healthy'' games. Because of the different categories and numerous user ratings, the assignment of games into ``dark'' and ``healthy'' groups is not binary but follows a gradient instead. Although neither term perfectly reflects all contained games, we will use them to stay in line with the website's use of the terms. %For clarity, we will adopt these labels when making references to these games to stay in line with our data. 

\subsection{Data Collection}
To extract the data, we developed a crawler using the scraping framework \href{https://docs.scrapy.org/en/latest/}{\textit{Scrapy}} to collect ratings, reported dark patterns as well as their absence from mobile games on the website. 
The initial data set comprised a total of $52,111$ games. Listed games stem from mobile app stores including Apple's App Store or Google's Play Store. In case a user of the website cannot find a game to rate it, they are asked to contact the website's host to add it. Overall, most games listed at the time of conducting this study did not contain any user ratings. After their removal, we could reduce the data set to 1,496 games with $85,388$ rated dark patterns, spanning $843$ ``dark'' and $653$ ``healthy'' games.

\subsection{Procedure}
To analyze this still relatively large data set, we opted for an exploratory data analysis~\cite{jebb_exploratory_2017}, a quantitative approach that allows the evaluation of complex data without requiring initial hypotheses. Instead, guiding questions help approach and navigate such data successfully. In our case, we formulated guiding questions (see Appendix~\ref{app:guiding-questions}) to account for the various variables between considered dark patterns across the four categories while focusing on possible relationships between them. This approach further allowed us to account for certain unknowns behind the data, as the website does not transparently state how user ratings are processed. 
To approach our research question, we developed a set of guiding questions for this data exploration (see Appendix \ref{app:guiding-questions}), aiming to understand the differences between ``healthy'' and ``dark'' games and how they are affected by certain dark patterns and their categories.
We processed the collected data, contained in a \texttt{json} file, using a Jupyter Notebook~\cite{jupyter_notebook} and the \texttt{Python} library \texttt{pandas} (version 2.0.3).

\section{Results}

\begin{table*}[t!]
\centering
\begin{tabular}{
    >{\raggedright\arraybackslash}p{2cm} 
    >{\centering\arraybackslash}p{1.2cm} 
    >{\centering\arraybackslash}p{1.2cm} 
    >{\centering\arraybackslash}p{1.2cm} 
    >{\centering\arraybackslash}p{1.2cm} 
    >{\centering\arraybackslash}p{1.2cm} 
    >{\centering\arraybackslash}p{1.2cm}}
    \toprule
    & \multicolumn{3}{c}{\textbf{Healthy Games (N=653)}} & \multicolumn{3}{c}{\textbf{Dark Games (N=843)}} \\ 
    \cmidrule(lr){2-4} \cmidrule(lr){5-7}
    \textbf{Category} & \textbf{Total} & \textbf{Average} & \textbf{SD} & \textbf{Total} & \textbf{Average} & \textbf{SD} \\ 
    \midrule
    \rowcolor[HTML]{E0E0E0} \textbf{Temporal} & 3167 & 4,85 & 23,82 & 18,630 & 22,09 & 69,67 \\
    \textbf{Monetary} & 2781 & 4,26 & 23,67 & 26,330 & 31,24 & 97,66 \\
    \rowcolor[HTML]{E0E0E0} \textbf{Social} & 990 & 1,52 & 10,42 & 11,835 & 14,04 & 55,10 \\
    \textbf{Psychological} & 3296 & 5,05 & 22,34 & 18,359 & 21,78 & 74,93 \\
    \midrule
    \textbf{Total} & 10,234 &  & & 75,154 &  & \\
    \bottomrule
\end{tabular}
\caption{This table presents an overview of the total number, average per game, and standard deviation of temporal, monetary, social, and psychological dark patterns for each ``healthy'' and ``dark'' game.}
\label{tab:dark-patterns-in-games}
\end{table*}

In this section, we report the results of our explorative data analysis. The section follows the different variables of the data and their relationship to ``healthy'' and ``dark'' games. To this end, we explored the presence of dark pattern categories in either group of games. We then dive into mobile games' revenue strategies and how they may be related to ``healthy'' and ``dark'' game groups. We began this analysis by conducting a normal distribution check using the Shapiro-Wilk tests for all variables. The test showed that the considered data are not normally distributed (p-values < .001).
Hence, we opted for using non-parametric statistical tests, mainly the Kruskal-Wallis test and the Chi-Square test of independence, for our further analysis.

\subsection{Dark Pattern Categories in Dark \& Healthy Games}
We explored the data focusing on dark pattern ratings for temporal, monetary, social, and psychological characteristics. Table~\ref{tab:dark-patterns-in-games} offers an overview of the total number of games per category that were included in our data set. Here, we follow our guiding questions further to explore individual aspects that play a role in making a game considerably ``dark'' or ``healthy''. %what makes a game ``healthy'' or ``dark'', look into these groups individually, consider a game's popularity, 

%\subsubsection{Differentiation within Various Dark Patterns}
%The first question, \textbf{"What differentiations exist within various dark patterns?"}, can be answered using the graph~\ref{fig:darkpatternpresent}. The data indicate that in Dark Games, Monetary dark patterns (\(N=26330\)) dominate, while Social dark patterns (\(N=11835\)) are the least represented. Psychological (\(N=18359\)) and Temporal dark patterns (\(N=18630\)) are present at roughly the same levels.
%To address the second question, \textbf{"When does a game qualify as Healthy and when as Dark, and how do they relate to each other?"}, we refer to Figures~\ref{fig:darkpatternpresent} and~\ref{fig:darkpatternabsence}. 
Between mobile games, users rated $85,388$ instances of dark patterns, with ``dark'' games containing a multitude of instances compared to ``healthy'' games. While the numbers presented in Table~\ref{tab:dark-patterns-in-games} allow for certain assumptions based on how games were rated, we wanted to statistically test which types of dark patterns are more likely to occur in which group of games, as either contains numerous instances of dark patterns.
As our data are not normally distributed, we conducted a Kruskal-Wallis test to compare the presence of dark pattern categories in ``dark'' and ``healthy'' games. For each category (temporal, monetary, social, and psychological), the test resulted in significant differences (p-values < .001), implying that ``dark'' games, indeed, contain more dark patterns of each category than ``healthy'' games. Notably, only $161$, or $10.76\%$, of the $1,496$ mobile games in our data (rated at least once) reported no instances of any dark patterns.
These differences between ``dark'' and ``healthy'' games may not be surprising as the data is structured to distinguish between games containing more or less dark patterns. Nonetheless, this analysis allows us to discuss the overall presence of dark patterns more precisely in the following analysis and our discussion.
%The conducted tests and generated statistics reveal that dark patterns appear more frequently in Dark Games (\(N=75154\)) than in Healthy Games (\(N=10234\)). Additionally, dark patterns are more often absent in Healthy Games (\(N=102157\)) compared to Dark Games (\(N=64792\)).
%\paragraph{Presence of Dark Patterns in Healthy Games}
%In response to the third question, \textbf{"Are Healthy Games completely devoid of dark patterns?"}, we examine Figure~\ref{tab:healthyGamesPresence}. This figure clearly shows that dark patterns are reported in the dataset of Healthy Games. Among these, Social dark patterns are the least reported (\(N=990\)). Temporal (\(N=3167\)), Monetary (\(N=2781\)), and Psychological dark patterns (\(N=3296\)) show similar frequencies, with Psychological dark patterns being the most common.

% \subsection{Price Analysis}
% \begin{figure*}[t!]
%  \centering
%     \begin{minipage}{.5\textwidth}
%         \centering
%         \includegraphics[width=1\textwidth]{Figures/VergleichPrice.pdf}
%         \caption{Another figure}
%         \label{fig:Price}
%     \end{minipage}%
%     \hfill
%     \begin{minipage}{.5\textwidth}
%         \centering
%         \includegraphics[width=1\linewidth]{Figures/VergleichInAppPurchases.pdf}
%         \caption{Another figure}
%         \label{fig:In App Purchases}
%     \end{minipage}
% \end{figure*}

%the characteristics of games falling into the "healthy" and "dark" categories, each distinguished by their attributes of "Price," "Ads," and "In-App Purchases." 
%\subsubsection{Free vs. Paid Games}
\subsection{Revenue Analysis}
\begin{figure*}[t!]
    \centering
    \includegraphics[width=1\textwidth]{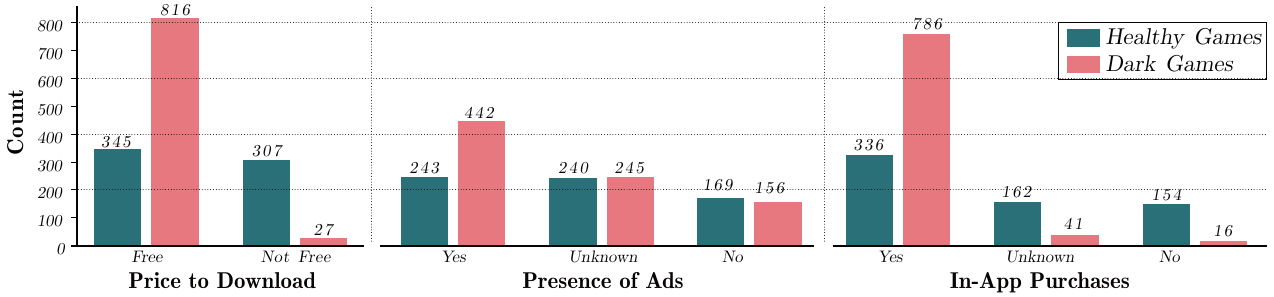}
    \caption{This figure presents bar charts for the revenue analysis of a game's price, contained advertisement, and in-app purchases between ``dark'' and ``healthy'' games.}
    \label{fig:bar-revenue}
    \Description{
    The figure presents three grouped bar charts comparing "Healthy Games" and "Dark Games" across three categories: Price to Download, Presence of Ads, and In-App Purchases. Each bar represents the count of games that fall under a specific criterion.

    Price to Download:
        Free: "Healthy Games" (345) vs. "Dark Games" (816)
        Not Free: "Healthy Games" (307) vs. "Dark Games" (27)

    Presence of Ads:
        Yes: "Healthy Games" (243) vs. "Dark Games" (442)
        Unknown: "Healthy Games" (240) vs. "Dark Games" (245)
        No: "Healthy Games" (169) vs. "Dark Games" (156)

    In-App Purchases:
        Yes: "Healthy Games" (336) vs. "Dark Games" (786)
        Unknown: "Healthy Games" (162) vs. "Dark Games" (41)
        No: "Healthy Games" (154) vs. "Dark Games" (16)
    }
\end{figure*}

To understand how the revenue of mobile games relates to included dark patterns, we considered data points for a game's price, used advertisement, and in-app purchases (see Figure~\ref{fig:bar-revenue}. Here, we present the distribution within each category and describe the relationships between these variables and the categories of games.

\paragraph{Pricing}
We considered whether mobile games had to be purchased or were offered as free-to-play before downloading. This does not include in-app purchases, which we compared later. %~\ref{fig:Price} depicts the cost distribution between the categories of ``healthy'' and ``dark'' games.
Within the ``healthy'' games category, games were almost evenly spread between free-to-play games $345$ ($53.0\%$) and $307$ ($47.0\%$) that cost some amount of money to purchase. Conversely, among ``dark'' games, $816$ ($96.8\%$) games were free-to-play, and only $27$ ($3.2\%$) games were not.
%The distribution of pricing in ``dark'' games reveals that the majority (\(96.80\%\)) of apps are labelled as "Free," while only a small fraction (\(3.20\%\)) are classified as "Not Free." In Healthy Games, approximately \(52.96\%\) of the games are labeled as "Free," while the remaining \(47.04\%\) are classified as "Not Free."
%With 816 of the total of 843 (96.80\%) ``dark'' games, this category contained significantly more free games compared to ``healthy'' games, with 345 that were free. With regard to paid games, we noticed an opposing trend: ``healthy'' games included 307 apps that had to be purchased, whereas ``dark'' games contained 27.
We conducted a Chi-Square test of independence to investigate the relationship between ``dark'' and ``healthy'' game categories with their pricing model to download the game. The analysis revealed a significant association between these variables (\(\chi^2 = 405.54\), \(df = 1\), \(p < .001\)). Hence, ``dark'' games are more likely to be free-to-play than ``healthy games''.
%The expected frequencies under the assumption of independence of variables were also significantly different (\(p < .001\)).

\paragraph{Advertisement}
%\paragraph{Presence of Advertisements}
Further, we wanted to understand the specific role advertisement plays between the categories of games.
Within ``healthy'' games, we recorded $243$ ($37.3\%$) that contained advertisements, with another $240$ ($36.8\%$) games having an ``unknown'' status, meaning that users made no clear statements regarding included advertisements. Finally, $169$ ($25.9\%$) games were marked to not contain advertisements. In contrast, ``dark'' games listed $442$ ($52.4\%$) games with advertisements, $245$ ($29.1\%$) games categorized as ``unknown'',  and $156$ ($18.5\%$) games without advertisements.
%The distribution of Ads in ``dark'' games shows that approximately  of the ``dark'' games contain some form of advertisements, while it is unknown for about \(34.16\%\) of the games and about \(4.34\%\) have no advertisements. In Healthy Games, approximately \(46.95\%\) display advertisements, \(46.58\%\) are unknown, and about \(6.47\%\) have no advertisements.
Again, we used a Chi-Square test of independence to explore how the presence of ads relates to the ``dark'' and ``healthy'' game categories. The test showed a significant relationship between these variables (\(\chi^2 = 34.6\), \(df = 2\), \(p < .001\)). This means that significantly more ``dark'' games contained advertisements.
%For ``dark'' games, there were 442 cases of ads, 245 unknown cases, and 156 cases with no ads, while in Healthy Games, there were 243 cases of ads, 240 unknown cases, and 169 cases with no ads. The expected frequencies under the assumption of independence of variables were also significantly different (\(p < .001\)).
%We visualized the results in Figure~\ref{fig:Ads}, which depicts the differences in the distribution of ads between Healthy Games and Dark Games.

\paragraph{In-App Purchases}
%\subsubsection{Presence of In-App Purchases}
With regards to in-app purchases,  $336$ ($54.0\%$) of all ``healthy'' games included features, while $162$ ($26.0\%$) games had an ``unknown'' status (indicating users who rated a game did not indicate the presence of in-app purchases), and 154 ($20.0\%$) games did not contain in-app purchases. In comparison, ``dark'' games had $786$ ($93.6\%$) games with in-app purchases, $41$ ($5.1\%$) games with an unclear status, and only $16$ ($1.3\%$) games not containing in-app purchases.
%The distribution of in-app purchases in dark games reveals that the majority (\(93.59\%\)) of apps have in-app purchases, while a small fraction (\(5.12\%\)) is classified as "Unknown," and an even smaller fraction (\(1.29\%\)) as "No." In Healthy Games, approximately \(53.98\%\) have in-app purchases, \(26.00\%\) are unknown, and \(20.02\%\) do not have in-app purchases.
A Chi-Square test shows a significant association between the presence of in-app purchases and ``healthy'' and ``dark'' game categories (\(\chi^2 = 345.87\), \(df = 2\), \(p < .001\)). %In Dark Games, there were 786 cases of in-app purchases, 41 unknown cases, and 16 cases with no in-app purchases, while in Healthy Games, there were 336 cases of in-app purchases, 162 unknown cases, and 154 cases with no in-app purchases. The expected frequencies under the assumption of independence of variables were also significantly different (\(p < .001\)).
``Dark'' games contained significantly more in-app purchase features than ``healthy'' games.

\section{Discussion}
In this study, we explored a large, user-generated data set to investigate how dark patterns manifest in $1,496$ mobile games based on the website \href{http://www.darkpatterns.games/}{www.darkpatterns.games}, utilizing dark patterns based on \citet{zagal_dark_2013}. To answer our research question, we studied how dark patterns occur between ``dark'' and ``healthy'' games as well as revenue-related factors. Here, we discuss the implications of our analysis while reflecting on related works. We begin by discussing our data source and the opportunities of community-based approaches and then continue answering our research question.
%two relatively large user-generated datasets to seek answers to several key questions. Users reported a total of ($N=75154$) instances of dark pattern in Dark Games and ($N=10234$) instances in Healthy Games [as of 19.05.2024]. But what does this mean?

%A direct comparison of dark patterns contained in either ``dark'' or ``healthy'' categories shows that different types of dark patterns were more prominent than others, affecting their players differently. The presence of temporal dark patterns in ``healthy'' games, for instance, could suggest that developers aimed to encourage players to spend more time in the game or repeat certain actions. 

%On one hand, this can have lead to negative and unhealthy player behaviour but, if done carefully, could also indicate that developers aimed to engage players long-term and provide a more immersive experience. In contrast, Monetary dark pattern in Dark Games aim to prompt players to spend money on in-app purchases or other paid features, highlighting a focus on immediate financial gains.

%Target Audience Approaches: The choice of patterns in the different categories may also indicate that developers of Healthy Games focus more on player needs and engagement, while Dark Games emphasize financial gains. This suggests that Healthy Games are designed for long-term player retention and a positive user experience, whereas Dark Games may be more focused on short-term financial returns, reflecting different target audience strategies.

%%%%%%%%%%%%%
%%%%%%%%%%%%%
\subsection{User Ratings of Dark Patterns in Mobile Games}
The source of our data in the form of a review website demonstrates the immense value of active online communities to voice concerns and unveil unethical designs, here with a specific focus on mobile games. In a similar vein, \citet{gray_what_2020} analysed member's posts of the \href{https://www.reddit.com/r/assholedesign/}{Reddit} community \textit{/r/assholedesign} to describe attributes of ``asshole''\footnote{We acknowledge the harshness of the term as an insult but wanted to stay in line to the original research by \citet{gray_what_2020} who defined related characteristics of practitioners.} practitioners who purposefully trick their users. These often large online communities spearhead efforts to generate data samples that often exceed restrictions of research funding alone. Ultimately, the collection of such data by the users themselves can be considered a form of participatory research. Moreover, these communities present research like Gray et al.'s and ours with opportunities to raise these concerns to the academic discourse, which has been shown to synergize with regulatory work~\cite{gray_dark_2021, gray_ontology_2024} and thus positively impact future experiences.  

However, it is important to consider the quality of such efforts. In this regard, the general ability of people to recognize dark patterns -- or differentiate between designs that contain instances and those that do not -- has been demonstrated in various studies (i.e.\cite{maier_dark_2020, bongard-blanchy_i_2021, di_geronimo_ui_2020, mildner_defending_2023}). Another thing these studies have in common, however, is that most participants who were able to recognize dark patterns were not sufficiently equipped to avoid their harmful effects. With regard to our work, these findings have at least two implications: First, although we cannot verify the origin of our data, we can assume that the community is unlikely to perform significantly differently compared to studies involving meaningful participant counts like, for example, \citet{bongard-blanchy_i_2021} with a total of $406$ and \citet{di_geronimo_ui_2020} with $589$ participants. Second, there is a chance that the actual number of contained dark patterns in the considered mobile games actually exceeds what the users have rated. This would align with the measured difficulties users face in successfully recognizing all present instances between these studies.

%%%%%%%%%%%%%
%%%%%%%%%%%%%
\subsection{Answering Our Research Question}
While exploratory in nature, our study revealed over $50k$ dark pattern instances between $1,496$ mobile games, offering answers to our research question. To this end, this research fills a gap within related work through its consideration of a large, quantitative data set to support existing qualitative findings from \citet{zagal_dark_2013} as well as later work~\citet{hadan_motivating_2024} building on their research.
One of the most critical findings of this study is the overall amount of dark patterns existing within games -- distributed between temporal, monetary, social, and psychological aspects, but also ``dark'' and ``healthy'' games (see Table~\ref{tab:dark-patterns-in-games}). While we have no insight into how ``dark'' and ``healthy'' labels were applied, we noticed that even games that are labelled as ``healthy'' contained dark patterns. It seems that the website's assignment of either label is based on a threshold that leverages the presence and absence of dark patterns with equal weights. Thus, the label's naming may cause problems as a ``healthy'' game may give people the false illusion that it is free of any dark patterns.  Although considering dark patterns within a continuum aligns with work by \citet{mildner_defending_2023}, our results revealed that only about $10\%$ of the mobile games observed showed no reported instances of dark patterns, implying that most players will face multiple instances of dark patterns, with potentially harmful implications on their well-being, which negatively affects trust and satisfaction with the games.

Finally, our data shows the prevalence of dark pattern types deployed between games. Overall, we noticed that social dark patterns, which leverage social elements or use social engineering to exploit a player's relationships with their friends and family, were perceived the least. And while psychological dark patterns (for instance, exploiting certain biases) counted the most within ``healthy'' games, monetary dark patterns were most prominent among ``dark games''. While these direct comparisons between ``dark'' and ``healthy'' games should be taken tentatively, particularly for the aforementioned lack of transparency of how they are determined, our results foreshadow which strategies are most dangerous for players. Our revenue-based analysis shows that free-to-play and ``freemium'' games feature significantly more dark patterns than their alternatives. Of course, developers should be rewarded for their work. However, the abundant presence of dark patterns throughout mobile games has serious implications for the well-being of players. Particularly, financial well-being and equity-related implications can increase burdens on people who may not be able to purchase a game at a higher price but will then be locked into pricing schemes, ultimately leading to higher costs and financial harm.

%Our findings raise questions about the ethics of free-to-play models and how they may be intertwined with practices designed to exploit users through in-app purchases or advertisements. While downloading a game for free might seem appealing, users could end up spending far more than they intended if key features or progress are locked behind paywalls. 

While further work is needed to better understand the factors that contribute to the broadness of dark patterns in games, the concerning outcomes of this study indicate that the gaming industry values greed and profit-based incentives more than their players' well-being. Within the field of HCI, different ethical design guidelines have been developed to mitigate harmful practice, for example, value-centred design~\cite{doorn_value_2013}, ethical design complexity \cite{chivukula_wrangling_2023}, or responsible design~\cite{mildner_mitigating_2024}. However, where such guidelines fail to ensure user safety, stricter regulations should safeguard players instead.

\section{Future Work \& Limitations}
Although we provide important insights into dark patterns in mobile games, we recognize the need for further investigation to achieve a more comprehensive understanding of the topic. We propose potential solutions and discuss integrating our findings into future research directions. One key limitation of our study is the lack of definitive data on the actual number of users visiting the website. Since the process of account creation is neither time-consuming nor influenced by many different factors, this could introduce bias, for instance in the nature of repeated entries, into our data, skewing the results. Because of (important) data protection regulations, there was no way to assess the origin of entries to confirm the quality of the data. Further, we examine only a subset of known dark patterns, as the website primarily relies on the framework defined by Zagal et al.\citep{zagal_dark_2013}. However, a recently published ontology for dark patterns~\cite{gray_ontology_2024} includes $64$ types, falling under high-, meso-, and low-level categories, and was derived from an initial set of $245$ instances across various academic and legal works. While focusing on context-specific dark patterns makes sense for a domain like mobile games, future research could explore high-level, domain-agnostic strategies that recur in different contexts.

%%%%%
Beyond the dark pattern discourse, but not covered in our data, we believe it would be valuable to investigate how game categories, target audiences, and other factors correlate with deployed dark patterns. With such knowledge, we could not only issue warnings for specifically at-risk populations but also promote and develop more effective countermeasures.
%Nevertheless, assuming the data has been generated by a significant number of users, it constitutes a dataset that we can utilize in future research. This dataset can serve as a foundation for continued investigation based on the findings of this study.

The community-based approach could also spark further future work by involving affected users. In this regard, \citet{aagaard_game_2022} proposes the introduction of a notification in app stores that indicates whether and which dark patterns are present in a game. Integrating such a badge could provide players with transparent information about the dark pattern present in a game before downloading it.
A promising extension of this idea could involve incorporating user-generated content. By engaging the community in this process, players could share their experiences with dark patterns, contributing to their identification and labeling. This approach would not only cover a wider variety of games but also raise awareness about dark patterns.

%The dataset e could be valuable in this context, as it already includes substantial user data. It is also crucial to examine whether the data from various digital distribution platforms are consistent, as the game offerings and their dark pattern might vary across platforms.
%In summary, both the introduction of a dark pattern Badge and the integration of user-generated content represent promising approaches to deepen the understanding of dark pattern in mobile games and enhance their transparency. Future research should focus on collecting additional data and examining the impact of implementing the dark pattern Badge on the gaming experience and player protection against undesirable influences. It is also important to explore the user perspective, investigating how players perceive the badge and whether it influences their game selection decisions.

%Overall, these approaches offer a promising foundation for future research to further investigate the phenomenon of dark pattern in games and develop potential solutions to protect players.
%%%%%%%%%%%%%%%%%%%%%%%%%%%%%
%        CONCLUSION         %
%%%%%%%%%%%%%%%%%%%%%%%%%%%%%
\section{Conclusion}
%This study investigates dark patterns in mobile games through an exploratory data analysis, revealing that even ``healthy'' games often contain manipulative design elements. By analyzing user-generated data, we emphasize the need for greater awareness and ethical design practices in the gaming industry. Future work should focus on expanding the understanding of dark patterns and fostering collaboration among stakeholders to promote fairer, healthier gaming experiences.
This study highlights the widespread presence of dark patterns in mobile games. Based on an explorative analysis of user-generated data, we demonstrate that manipulative design practices are embedded across a sample of $1,496$ mobile games, raising ethical concerns with a focus on revenue models and different types of dark patterns. Between temporal, monetary, social, and psychological dark patterns, the data reveals over $85,000$ instances of dark patterns trying to subvert users' choices against their best interest. Our research underscores the importance of community-based efforts to mitigate the negative effects of dark patterns, promoting fairer and healthier gaming experiences for all players.

\begin{acks}
    We thank the host of the website \textit{www.darkpattern.games} for allowing us to crawl and use the data in this research.
    This work was partially supported by the Leibniz ScienceCampus Bremen Digital Public Health, which is jointly funded by the Leibniz Association (W72/2022), the Federal State of Bremen, and the Leibniz Institute for Prevention Research and Epidemiology – BIPS.
\end{acks}

\bibliographystyle{ACM-Reference-Format}
\bibliography{references.bib}

\appendix
\section{Dark Pattern Categories from darkpatterns.games}
\label{app:dp-games-categories}
The website allowed users to rate mobile games based on four categories, including four specific types of dark patterns each. Table~\ref{tab:dark_patterns} offers a full overview in this regard.

% \begin{enumerate}
%     \item Temporal Dark Patterns
%     \begin{itemize}
%         \item Playing by Appointment
%         \item Daily Rewards
%         \item Grinding
%         \item Advertisements
%         \item Infinite Treadmill
%         \item Can't Pause or Save
%         \item Wait To Play
%     \end{itemize}
%     \item Monetary Dark Patterns
%     \begin{itemize}
%         \item Pay to Skip
%         \item Premium Currency
%         \item Pay to Win
%         \item Artificial Scarcity
%         \item Accidental Purchases
%         \item Recurring Fee
%         \item Gambling / Loot Boxes
%         \item Power Creep
%         \item Pay Wall
%         \item Waste Aversion
%         \item Anchoring Tricks
%     \end{itemize}
%     \item Social Dark Patterns
%     \begin{itemize}
%         \item Social Pyramid Scheme
%         \item Social Obligation / Guilds
%         \item Friend Spam / Impersonation
%         \item Reciprocity
%         \item Encourages Anti-Social Behavior
%         \item Fear of Missing Out
%         \item Competition
%     \end{itemize}
%     \item Psychological Dark Patterns
%     \begin{itemize}
%         \item Invested / Endowed Value
%         \item Badges / Endowed Progress
%         \item Complete the Collection
%         \item Illusion of Control
%         \item Variable Rewards
%         \item Aesthetic Manipulations 
%         \item Optimism and Frequency Biases
%     \end{itemize}
% \end{enumerate}

\begin{table}[h!]
    \centering
    \begin{tabular}{ll}
        \toprule
        \textbf{Category} & \textbf{Dark Patterns} \\
        \midrule
        Temporal & Playing by Appointment \\
                 & Daily Rewards \\
                 & Grinding \\
                 & Advertisements \\
                 & Infinite Treadmill \\
                 & Can't Pause or Save \\
                 & Wait To Play \\
        \midrule
        Monetary & Pay to Skip \\
                 & Premium Currency \\
                 & Pay to Win \\
                 & Artificial Scarcity \\
                 & Accidental Purchases \\
                 & Recurring Fee \\
                 & Gambling / Loot Boxes \\
                 & Power Creep \\
                 & Pay Wall \\
                 & Waste Aversion \\
                 & Anchoring Tricks \\
        \midrule
        Social & Social Pyramid Scheme \\
               & Social Obligation / Guilds \\
               & Friend Spam / Impersonation \\
               & Reciprocity \\
               & Encourages Anti-Social Behavior \\
               & Fear of Missing Out \\
               & Competition \\
        \midrule
        Psychological & Invested / Endowed Value \\
                      & Badges / Endowed Progress \\
                      & Complete the Collection \\
                      & Illusion of Control \\
                      & Variable Rewards \\
                      & Aesthetic Manipulations \\
                      & Optimism and Frequency Biases \\
        \bottomrule
    \end{tabular}
    \caption{Dark Patterns by Category. Based on \href{https://www.darkpattern.games/}{darkpattern.games}.}
    \label{tab:dark_patterns}
\end{table}

\section{Guiding Questions for the Explorative Data Analysis} \label{app:guiding-questions}
\begin{enumerate}
\item What differentiation exists between various dark pattern types?
\item When does a game qualify as ``healthy'', and when does it transition into ``dark'' territory, and how do they relate to each other?
\item Are ``healthy'' games completely devoid of dark patterns?
\item Which category is the most popular and boasts a larger user base?
\item Is payment always necessary to access dark pattern-free games?
\item  How do the frequency and nature of dark patterns differ between the two game groups?
\item  Are there specific dark patterns that are more prevalent in ``dark'' games compared to ``healthy'' games?
\end{enumerate}

\end{document}